\documentclass[showpacs,amsmath,amssymb,aps,prd]{revtex4}
\input{epsf}
\usepackage{graphicx}
\usepackage{natbib}
\def\X{{\mathrm{x}}}
\def\Y{{\mathrm{y}}}
\def\Z{{\mathrm{z}}}

\def\n{{\rm n}}
\def\p{{\rm p}}
\def\e{{\rm e}}
\def\s{{\rm s}}
\def\c{{\rm c}}

\def\be{\begin{equation}}
\def\ee{\end{equation}}
\def\bea{\begin{eqnarray}}
\def\eea{\end{eqnarray}}

\begin{document}

\title{A Flux-Conservative Formalism for Convective and Dissipative 
Multi-Fluid Systems, with Application to Newtonian Superfluid Neutron 
Stars}
\author{N.~Andersson}
\email{na@maths.soton.ac.uk}
\affiliation{School of Mathematics, University of Southampton, 
Southampton SO17 1BJ, United Kingdom} 
\author{G.~L.~Comer}
\email{comergl@slu.edu}
\affiliation{Department of Physics \& Center for Fluids at All Scales, 
Saint Louis University, St.~Louis,
MO, 63156-0907, USA}

\date{\today}

\begin{abstract}
We develop a flux-conservative formalism for a Newtonian multi-fluid 
system, including dissipation and entrainment (i.e.~allowing the momentum 
of one fluid to be a linear combination of the velocities of all fluids). 
\ Maximum use is made of mass, energy, and linear and angular momentum 
conservation to specify the equations of motion.\ Also used extensively 
are insights gleaned from a convective variational action principle, key 
being the distinction between each velocity and its canonically conjugate 
momentum (which is modified because of entrainment). \ Dissipation is 
incorporated to second order in the ``thermodynamic forces'' via the 
approach pioneered by Onsager, which makes it transparent how to guarantee 
the law of increase of entropy. \ An immediate goal of the investigation 
is to understand better the number, and form, of independent dissipation 
terms required for a consistent set of equations of motion in the 
multi-fluid context. \ A significant, but seemingly innocuous detail, is 
that one must be careful to isolate ``forces'' that can be written as 
total gradients, otherwise errors can be made in relating the net 
internal force to the net externally applied force. \ Our long-range aim 
is to provide a formalism that can be used to model dynamical multi-fluid 
systems both perturbatively and via fully nonlinear 3D numerical 
evolutions. \ To elucidate the formalism we consider the standard model 
for a heat-conducting, superfluid neutron star, which is believed to be 
dominated by superfluid neutrons, superconducting protons, and a highly 
degenerate, ultra-relativistic gas of normal fluid electrons. \ We 
determine that in this case there are, in principle, $19$ dissipation 
coefficients in the final set of equations. \ A final reduction of the 
system is made by neglecting heat conduction. \ This leads to an 
extension of the standard two-fluid model for neutron star cores, which 
has been used in a number of previous applications, and illustrates how 
mutual friction is represented in our formalism. 
\end{abstract}

\maketitle

\section{Introduction}

The chemistry community has great experience in dealing with multi-fluid 
systems. \ Perhaps indicative of the cultural divide between research 
areas is the development of a multi-fluid literature for superfluid 
neutron stars which largely ignores the chemistry successes. \ 
Of particular note is the Onsager formulation of dissipative, multi-fluid 
systems \cite{onsager31:_symmetry}. \ On the other hand, action-based 
derivations of the equations of motion for multi-fluid systems 
\cite{prix04:_multi_fluid,carter04:_cov1,carter03:_cov2,carter04:_cov3} 
have, as far as we can tell, not crossed over in the other direction. \ 
In this work, we  attempt to span the divide by utilizing fully mass, 
energy, and linear and angular momentum conservation, the Onsager 
formulation, and action-based constructs to build a general dissipative 
model for multi-fluid systems. We facilitate our use of the conservation 
laws by borrowing from the engineering community the idea of ``control'' 
volumes. \ Even though we restrict the discussion to the Newtonian 
regime, we do not hesitate to use ideas that have been developed for 
general relativity. \ In fact, the analysis is often much simplified if 
the true covariant nature of the physics is retained. 

Although much of our analysis is general, and can be used in any 
multi-constituent fluid context---for example, superfluid Helium and 
bubbly liquid/gas mixtures (see \cite{drew} for a general review of 
two-phase flow, and a series of papers by Geurst 
\cite{geurst86,geurst88a,geurst88b} for interesting discussions), our 
immediate goal is to produce a formalism that can be applied to Newtonian 
superfluid neutron stars, in order to  model their detailed dynamics. \ 
The importance of this target is enhanced by the fact that the first 
generation of large-scale, ground-based, gravitational-wave detectors are 
now operating near their design sensitivity \cite{LIGO05PRL}. \ Mature 
neutron stars can radiate gravitationally in a number of ways, many of 
which depend crucially on dissipation in the superfluid interior. \ An 
obvious example is the potentially strong damping of gravitational-wave 
driven mode-instabilities (eg.~of the so-called r-modes) due to either 
superfluid mutual friction 
\cite{Andersson:2002ch,lindblom00:_r_modes_superfl_neutr_stars}  
or a viscous boundary layer at the neutron star core-crust interface 
\cite{rieutord01:_ekman,glampedakis04:_boundary}. \ Other astrophysical 
contexts requiring a detailed model for superfluid dissipation include 
the pulsar glitches \cite{lyne}, the standard model for which is based on 
the transfer of angular momentum between components rotating relative to 
each other, and possible neutron star free precession \cite{ianj}, the 
damping of which depends crucially on the interior viscosity. 

In the standard superfluid neutron star model, the neutrons in the inner 
crust and the outer core are believed to be superfluid, and the protons 
in the outer core are thought to be superconducting. \ Because of overall 
charge neutrality, the neutrons and protons will coexist with a normal, 
but highly degenerate fluid of electrons (and possibly muons in the more 
dense regions). \ At the core temperature expected in a mature neutron 
star (likely four orders of magnitude below the relevant Fermi 
temperatures), the superfluid and superconducting nature, respectively, 
of the neutrons and protons implies that they function as dynamically 
independent, interpenetrating fluids. \ The conventional wisdom is that 
viscous damping of the oscillations of these objects is a result of 
electron-electron scattering. \ However, this begs the following 
question: how is it that damping of the electron motion on the one hand 
leads to damping, on the other hand, of the superfluid neutrons, which 
make up the bulk of the matter in the star? \ Despite these kinds of 
problems having been studied for decades 
\cite{alpar84:_rapid_postglitch,mendell91:_diss}, 
the answer is not at all obvious from the extant literature (although we 
have recently clarified some aspects in 
\cite{andersson04:_viscosity_coeff,mutualf}). \ Indeed, the conventional 
wisdom seems to have been guided by the classic discussion of Cutler and 
Lindblom \cite{cutler87:_viscosity}, who noted that ``neutron star matter 
becomes more viscous in the superfluid state than it was in the normal 
state.'' \ However, while they included the effect of superfluidity in 
their determination of the viscosity coefficients, they did not consider 
the effect on the actual fluid dynamics. \ Obviously aware of this fact, 
they remark that ``The superfluidity of neutron star matter, will, of 
course, have drastic effects both on the dissipation coefficients and on 
the dynamics of this material.'' \ The implications of the first 
statement on the possibility of detectable gravitational radiation from 
neutron stars are potentially far-ranging, leading most directly to the 
conclusion that gravitational-wave driven instabilities are ineffective 
in mature neutron stars \cite{Andersson:2002ch}, provided, of course, 
that superfluid dynamics can be ignored. \ However, subsequent work on 
the role of superfluidity in neutron stars reveals that the effects on 
the dynamics can be significant, and sometimes unanticipated (see 
\cite{comer02:_zero_freq_subspace,comer03:_review,acp03:_twostream_prl,andersson04:_twostream} 
for recent discussions). 

To date, most studies of superfluid neutron stars have drawn upon 
Landau's original two-fluid model for He$^4$ \cite{landau59:_fluid_mech}. 
\ In the case of neutron stars, both neutrons and protons are expected
to form Cooper pairs, which then form condensates which at the fluid level
can be described by the standard equations. \ The main distinction in the 
neutron star case is that the two fluids no longer represent the 
superfluid and normal parts of a single particle species. \ Instead, the 
two degrees of freedom describe the neutrons and a conglomerate of all 
charged components which are expected to be electromagnetically coupled 
on a short timescale. \ Following the traditional route, Mendell 
\cite{mendell91:_nondiss,mendell91:_diss} extended the non-dissipative 
zero-temperature equations to account for the main dissipation 
mechanisms. \ The result is a set of equations which include the mutual 
friction coupling due to electrons scattering off of rotational vortices 
in the condensates (see also \cite{mutualf}). \ These equations were 
later used to study the damping of unstable modes of oscillation in a 
spinning neutron star. \ The conclusion drawn from this work is that 
mutual friction is a key dissipation mechanism which may, in fact, 
suppress rotational instabilities entirely 
\cite{lindblom95:_does_gravit_radiat_limit_angul,lindblom00:_r_modes_superfl_neutr_stars}.

As the work of Mendell shows, the mathematical description of the problem 
becomes considerably more complicated once the so-called entrainment 
effect \cite{andreev75:_three_velocity_hydro}, which 
accounts for the fact that the flow of one fluid may impart 
momentum in the other fluid, is included. \ This complexity is partly due 
to a common confusion concerning Landau's description of superfluid 
hydrodynamics. \ At the heart of the problem lies a failure to 
distinguish clearly between transport velocities and momenta. \ This 
distinction is made clear in recent variational formulations of the 
problem 
\cite{prix04:_multi_fluid,carter04:_cov1,carter03:_cov2,carter04:_cov3}, 
work which draws heavily on the fully covariant relativistic description 
of Carter and collaborators  
\cite{carter89:_covar_theor_conduc,comer93:_hamil_multi_con,comer94:_hamil_sf,carter95:_kalb_ramond,carter98:_relat_supercond_superfl,langlois98:_differ_rotat_superfl_ns,prix00:_cov_vortex}, 
and which incorporates entrainment in a natural way. \ This leads to a 
mathematical description which is in many ways less complex than the 
standard one.  

Our present analysis differs somewhat from the recent variational 
derivations 
\cite{prix04:_multi_fluid,carter04:_cov1,carter03:_cov2,carter04:_cov3}
by emphasizing the conservations laws for mass, linear and angular 
momentum, and energy. \ We find this approach appealing because it stays 
close to our physical intuition. \ Furthermore, we believe that our 
discussion may serve as a useful ``introduction'' to the more technically 
involved work described in 
\cite{prix04:_multi_fluid,carter04:_cov1,carter03:_cov2,carter04:_cov3}. 
\ Of course, we rely heavily on the variational methods for rigorous 
mathematical support of our results in the non-dissipative case. \ 
Until very recently there had been no serious discussion of the 
dissipative problem within the variational framework. \ The discussion by 
Carter and Chamel 
\cite{carter04:_cov1,carter03:_cov2,carter04:_cov3} provided a useful 
first step by translating (and extending somewhat) the covariant 
relativistic model devised by Carter \cite{carter89:_covar_theor_conduc} 
and Carter and Khalatnikov \cite{carter94:_canon_formul_newton_superfl}, 
yet the discussion remains somewhat abstract. \ A key aim of the present 
discussion is to make contact with applications and prepare the ground 
for building realistic models of astrophysical superfluids. 

The present work has two main aims. \ First of all, we want to consider 
the multi-fluid dissipation problem in ``complete'' generality. \ In 
doing this we rely on the key conservation laws and ask what the most 
general form for the dissipation and interaction forces may be. \ The 
dissipative terms are then constrained by ensuring that the second law of 
thermodynamics is satisfied. \ To make progress in this direction we make 
use of Onsager's celebrated symmetry principle 
\cite{onsager31:_symmetry}. \ Our derivation follows the guidelines set 
out in work on superfluid ${\rm He}^4$ (see \cite{khalatnikov65:_introd} 
and \cite{putterman74:_superfluid} for nice descriptions) and is 
basically an extension of the general analysis championed, for instance, 
by Landau and Lifshitz \cite{landau59:_fluid_mech} for one-fluid 
systems. \ We extend previous work in two important ways: we include the 
entrainment effect in a (hopefully) clear way, and we also allow for the 
presence of mutual interaction terms which have traditionally not been 
considered (eg.~in \cite{carter04:_cov1,carter03:_cov2,carter04:_cov3}). 
\ The latter turns out to be crucial if we want to be able to represent 
the mutual friction forces which arise due to the presence of vortices. \ 
Up to this point, i.e.~in Secs.~\ref{fluxcon} and \ref{onsager}, the 
discussion is general and should be relevant for any multi-fluid system. 
\ Our second objective is met in Sec.~\ref{npee}, where we specify the 
equations to the conditions that should prevail in the outer core of a 
mature neutron star. \ We consider a mixture of superfluid neutrons, 
superconducting protons, normal electrons and entropy. \ This model is 
instructive in a number of ways. \ We show how the dynamical degrees of 
freedom are reduced in a natural way if we assume that the electrons and 
protons are coupled electromagnetically. \ (It should be noted that we do 
not account for detailed electromagnetic effects in this analysis 
\cite{prix_magnetic}.) \ We also deduce that this system in general 
requires the specification of no less than 19 distinct dissipation 
coefficients. \ This discussion is followed, and the paper concluded, by 
a final reduction to the dissipative extension of the standard two-fluid 
model which has been used in recent work on neutron star dynamics. \ This 
model, which is obtained by neglecting heat conduction, provides the 
simplest system that should be considered in ``realistic'' applications. 
\ We discuss the interpretation of the various dissipative terms, some of 
which appear to be new. \ In particular, we show how the standard form 
for the mutual friction force \cite{mutualf} is represented within our 
formalism.

\section{Conservation Laws for Multi-Fluid Systems} \label{fluxcon}

Any general fluid formalism must respect the three laws concerning
\begin{itemize}
    \item Conservation of Mass
    \item Conservation of Linear and Angular Momentum 
    \item Conservation of Energy
\end{itemize}
Although entropy is not conserved in general, the law of increase of 
entropy is also a vital property of any fluid. \ In imposing these 
conservation laws for a multi-fluid system, we consider some ``control 
volume'' $V$ with  boundary $\partial V$, letting $\eta_i$ denote the 
unit normal form to $\partial V$. \ The control volume is to be 
contrasted with the notion of a fluid element, which is a region of the 
fluid small enough to be considered ``infinitesimal'' with respect to the 
whole fluid and yet contain enough particles that a (local) thermodynamic 
treatment is warranted. 

We consider a general multi-fluid system, the dynamical description of 
which is based on determining the space and time dependence of each 
particle number density $n_\X$ and the corresponding particle number 
density current $n^i_\X$. \ These are natural variables to use in 
describing the system. \ It is assumed that the different fluids are 
interpenetrating, i.e.~the different constituents do not have to be  
separated by interfaces. \ We let $\X,\Y,\Z$ be particle constituent 
indices that range over the whole set of fluids in the system. \ Repeated 
constituent indices will not satisfy a summation convention, but vector 
indices $i,j,k$ do satisfy such a convention (as usual). \ Ostensibly, 
all formulas will be written in a general coordinate basis, with 
$g_{i j} = g_{j i}$ representing the (flat-space) metric, except for 
certain integrals written below, where a Cartesian basis is implicit 
(since tensor components are being integrated). \ The metric is used to 
raise and lower indices, e.g.~for a vector $v_i = g_{i j} v^j$ and 
$v^i = g^{i j} v_j$, where the inverse metric $g^{i j}$ satisfies 
$g^{j k} g_{k i} = \delta^j{}_i$ and $\delta^j{}_i$ is the Kronecker-delta 
tensor. \ When integrating over volumes, the measure is given by 
${\rm d} V = \sqrt{g} {\rm d}^3 x$ where $g$ is the determinant of the 
metric. \ Finally, all spatial derivatives are given by the covariant 
derivative $\nabla_i$ that is compatible with the metric 
(i.e.~$\nabla_i g_{j k} = 0$).

\subsection{Conservation of Mass}

Let $m^\X$ denote the ``particle'' mass of the $\X^{\rm th}$-constituent, 
so that $\rho_\X = m^\X n_\X$ is the mass density of that same 
constituent. \ The total (local) mass density of the system, $\rho$, is 
thus
\be
    \rho = \sum_\X \rho_\X = \sum_\X m^\X n_\X \ .
\ee
The total mass-density current $\rho^i$ is likewise given by
\be
    \rho^i = \sum_\X m^\X n^i_\X \ .
\ee
When $\rho$ is integrated over the control volume $V$, we obviously 
determine the mass of fluid ${\rm M}$ contained in $V$, i.e.
\be
    {\rm M} = \int_V \rho~{\rm d} V \ .
\ee
When $\rho^i$ is integrated over $\partial V$, we have a 
representation of the amount of mass that leaves (or enters, depending on 
the situation) $V$. \ Overall conservation of mass thus means
\be
   \frac{{\rm d}}{{\rm d} t} \int_V \rho~{\rm d} V = - \int_{\partial V} 
   \rho^i \eta_i~{\rm d} A = - \int_V \nabla_i \rho^i~{\rm d} V \ .
\ee
This implies the local condition 
\be
    \partial_t {\rho} + \nabla_i \rho^i = 0 \ . \label{mcon}
\ee

If $\Gamma_\X$ represents the particle number creation rate per unit 
volume of the $\X^{\rm th}$-constituent, then we have
\be
    \partial_t{n}_\X + \nabla_i n^i_\X = \Gamma_\X \ . \label{numx}
\ee
Multiplying this by $m^\X$, summing over all $\X$, and using the 
conservation of mass, Eq.~(\ref{mcon}), one immediately sees that 
\be
    \sum_\X m^\X \Gamma_\X = 0 \ . \label{pcreacon}
\ee
Worth noting at this point is that we will consider the entropy to be a 
fluid, the main distinction being that its mass is zero. \ Its ``number'' 
density is simply the entropy per unit volume, and it is assumed to 
satisfy a creation rule like Eq.~(\ref{numx}).

\subsection{Conservation of Linear Momentum}

Let $p^\X_i$ represent the (local) linear momentum per fluid element 
(which will later be understood to be canonically conjugate to $n^i_\X$) 
of the $\X^{\rm th}$-constituent. \ The total (local) linear momentum of 
the fluid $p_i$ will then obviously be given by
\be
    p_i = \sum_\X p^\X_i \ .  
\ee 
Likewise let $\pi_i$ denote the total linear momentum density, so that
\be
    \pi_i = \sum_\X n_\X p^\X_i \equiv \sum_\X \pi^\X_i \ .
\ee
The total (global) linear momentum ${\rm P}_i$ of the fluid contained in 
the control volume will be given by
\be
    {\rm P}_i = \int_V \pi_i~{\rm d} V \ . \label{totP}
\ee
Finally, let $T^j{}_i$ represent the $i^{\rm th}$-component of the total 
linear momentum that flows (per unit time and unit area) in a direction 
perpendicular to the $j^{\rm th}$-direction, and let $f_i$ be the net 
external force density acting on the control volume. \ Overall 
conservation of linear momentum in $V$ is then given by
\be
   \frac{{\rm dP}_i}{{\rm d}t} = \frac{{\rm d}}{{\rm d} t} \int_V 
   \pi_i~{\rm d} V = - \int_{\partial V} T^j{}_i \eta_j~{\rm d} A + 
   \int_V f_i {\rm d} V = - \int_V \left(\nabla_j T^j{}_i - f_i\right) 
   {\rm d} V \ , \label{momconv}
\ee
which implies that
\be
    \partial_t{\pi}{}_i + \nabla_j T^j{}_i = f_i \ . \label{momcon}
\ee
Here we could have added a term that corresponds to the net internal 
forces. \ However, the weak law of action and reaction implies that this 
contribution is zero \cite{goldstein80:_cm}. \ There are also the 
well-known difficulties for action and reaction when charged particles 
are present (cf.~\cite{goldstein80:_cm}).

The individual fluid components will not satisfy exactly the same 
conservation law, since momentum and energy can be exchanged between 
them. \ However, it is worthwhile to assume that the 
$\X^{\rm th}$-constituent satisfies an equation of the form
\be
    \partial_t{\pi}{}^\X_i + \nabla_j T^\X{}^j{}_i = \overline{f}{}^\X_i 
    \ , \label{eulerx}
\ee
noting that this is still completely general, as nothing has been 
proposed about either $T^\X{}^j{}_i$ or $\overline{f}{}^\X_i$. \ That 
being said, one should not be too surprised by such a choice, since it 
represents the limiting form familiar from single-fluid studies. \ 
Consistency with overall linear momentum conservation now demands that
\be
   f_i - \sum_{\X} \overline{f}{}^\X_i = \nabla_j \left(T^j{}_i - 
   \sum_{\X} T^\X{}^j{}_i\right) \ . \label{consist}
\ee
Although it may seem natural to suppose that the left- and right-hand 
sides should vanish independently, we will establish below that this is 
not correct.

\subsection{Conservation of Energy}

Let ${\cal U}$ denote the total energy per unit volume in $V$, and 
${\rm U}$ be the total energy of the control volume, i.e.
\be
    {\rm U} = \int_V {\cal U}~{\rm d} V \ . \label{toteng}
\ee
Also let $Q^i$ denote the flow of energy (per unit time and unit area) 
perpendicular to the $i^{\rm th}$-direction, and $\epsilon^{\rm ext}$ the 
energy creation rate per unit volume due to external sources. \ 
Conservation of energy in $V$ is then naturally expressed as 
\be
  \frac{\rm dU}{{\rm d} t} = \frac{{\rm d}}{{\rm d} t} \int_V 
  {\cal U}~{\rm d} V = - \int_{\partial V} Q^i \eta_i~{\rm d} A + \int_V 
  \epsilon^{\rm ext} {\rm d} V = - \int_V \left(\nabla_i Q^i - 
  \epsilon^{\rm ext}\right) {\rm d} V \ , \label{engconv}
\ee
or
\be
    \partial_t{{\cal U}} + \nabla_i Q^i = \epsilon^{\rm ext} \ . 
    \label{engcon}
\ee
Here it should be noted that we must specify a relation ${\cal U} = 
{\cal U}(n_\X,n^i_\X)$ to close the system of equations. \ That is, we 
need to provide an equation of state for the matter. 

\subsection{Conservation of Angular Momentum} 
\label{angmomcons}

In order to get an appropriate definition of the total angular 
momentum ${\rm L}^i$ in the control volume, let us first reconsider the 
total linear momentum ${\rm P}_i$  in the context of Newton's Second Law. 
\ We need to determine those internal and external contributions that act 
within the fluid to change ${\rm P}_i$, so as to properly define the net 
torque. 

Referring back to Eq.~(\ref{totP}), we can use ${\rm P}_i$ in Newton's 
Second Law to determine the net ``force'' ${\rm F}_i$ acting on the 
control volume:
\bea
    {\rm F}_i &\equiv& \frac{{\rm d}{\rm P}_i}{{\rm d} t} 
              = \int_V \partial_t \pi_i~{\rm d} V = \int_V \left(f_i - 
                \nabla_j T^j{}_i\right) {\rm d} V \ . \label{2ndlf}
\eea
This suggests that the appropriate definition of the total torque 
${\rm T}^i$ acting on the control volume (in Cartesian coordinates) is
\be
    {\rm T}^i \equiv \frac{1}{2} \epsilon^{i j k} \int_V \left(x_j 
                     \left[f_k - \nabla_l T^l{}_k\right] - x_k \left[f_j 
                     - \nabla_l T^l{}_j\right]\right)~{\rm d} V \ .
\ee
To see that this definition is consistent, we consider the time rate 
of change of ${\rm L}^i$ and show that it is equal to ${\rm T}^i$.

The total angular momentum of the control volume is defined to be
\be
    {\rm L}^i \equiv \frac{1}{2} \epsilon^{i j k} \int_V \left(x_j \pi_k 
                     - x_k \pi_j\right)~{\rm d} V \ ,
\ee
and its time rate of change is equal to
\bea
    \frac{{\rm d} {\rm L}^i}{{\rm d t}} &=& \frac{1}{2} \epsilon^{i j k} 
    \int_V \left(x_j \partial_t \pi_k - x_k \partial_t \pi_j\right)~{\rm 
    d} V \cr
     && \cr
    &=& \frac{1}{2} \epsilon^{i j k} \int_V \left(x_j \left[f_k - 
    \nabla_l T^l{}_k\right] - x_k \left[f_j - \nabla_l T^l{}_j\right]
    \right)~{\rm d} V \cr
     && \cr
    &=& {\rm T}^i \ , \label{Ldot}
\eea
which is the desired result. \ Of course, Eq.~(\ref{Ldot}) implies that 
the total angular momentum contained within the control volume is 
conserved when the net torque vanishes. \ We can use this to our 
advantage now, by showing that conservation results only if the symmetry 
$T_{i j} = T_{j i}$ holds. 

When a system is completely isolated, and the control volume is such that 
it completely contains the system and has a boundary taken well outside 
the system, we should find that ${\rm L}^i$ is conserved, i.e.~constant 
in time, if in addition the net force $f_i$ acting is zero. \ 
Eq.~(\ref{Ldot}) can be rewritten as
\be
    \frac{{\rm d} {\rm L}^i}{{\rm d t}} = \frac{1}{2} \epsilon^{i j k} 
    \left\{\int_V \left(x_j f_k - x_k f_j\right)~{\rm d} V - \int_V 
    \left(T_{j k} - T_{k j}\right)~{\rm d} V - \int_{\partial V} 
    \left(x_j T^l{}_k - x_k T^l{}_j\right) \eta_l~{\rm d} A\right\} \ . 
    \label{Ldot2}
\ee
The first term on the right vanishes because the net force $f_i$ is 
zero. \ The last term vanishes because we have taken the boundary of the 
control volume well outside the system. \ Hence, if angular momentum is 
to be conserved, the middle term must also vanish, which implies $T_{i j} 
= T_{j i}$.

Because of the inherent linearity in our definitions when summing over 
the constituent indices, we immediately see that an appropriate 
definition for the control volume force $F^\X_i$ acting on the 
x$^{\rm th}$-constituent is
\be
    {\rm F}^\X_i = \int_V \left(f^\X_i - \nabla_j T^\X{}^j{}_i\right) 
                   {\rm d} V \ , 
\ee
while for the control volume angular momentum we have
\be
    {\rm L}^i_\X = \frac{1}{2} \epsilon^{i j k} \int_V \left(x_j \pi^\X_k 
                   - x_k \pi^\X_j\right) {\rm d} V \ , 
\ee
where the corresponding torque is
\be
    {\rm T}^i_\X = \frac{1}{2} \epsilon^{i j k} \int_V \left(x_j \left[
                   \overline{f}{}^\X_k - \nabla_l T^\X{}^l{}_k\right] - 
                   x_k \left[\overline{f}{}^\X_j - \nabla_l T^\X{}^l{}_j
                   \right]\right) {\rm d} V \ . 
\ee
This naturally implies
\be
    \frac{{\rm d} {\rm L}^i_\X}{{\rm d t}} = \frac{1}{2} \epsilon^{i j k} 
    \left\{\int_V \left(x_j \overline{f}{}^\X_k - x_k \overline{f}{}^\X_j
    \right)~{\rm d} V - \int_V \left(T^\X{}_{j k} - T^\X{}_{k j}
    \right)~{\rm d} V - \int_{\partial V} \left(x_j T^\X{}^l{}_k - x_k 
    T^\X{}^l{}_j\right) \eta_l~{\rm d} A\right\} \ .
\label{ldd}\ee
At this point we make an important observation: In contrast to the 
comments following Eq.~(\ref{Ldot2}), we argue that it is a mistake to 
infer that $T^\X{}_{i j} = T^\X{}_{j i}$. \ This assumption would be too 
restrictive. \ In particular, we will show later that it would not allow 
us to represent  the standard form for the vortex-mediated mutual 
friction in a superfluid system \cite{mutualf}.

\subsection{Lagrange's Generalized Action Principle and Equations of 
Motion}

Further progress  can be made in specifying the unknown quantities 
${\cal U}$, $Q^i$, $\overline{f}{}^\X_i$, $T^\X{}^j{}_i$ and $T^j{}_i$, 
assuming that $f_i$ and $\epsilon^{\rm ext}$ are prescribed. \ Consider 
that there is a Lagrangian density ${\cal L}$ whose variation leads to 
the multi-fluid equations in the fully conservative, non-dissipative, 
case. \ It consists of the total kinetic energy density minus the 
``potential'' energy density, which we take to be the total internal 
energy density ${\cal E}(n_\X,n^i_\X)$, 
\be
   {\cal L} = \sum_\X \frac{1}{2} m^\X g_{i j} n^i_\X n^j_\X / n_\X 
              - {\cal E}(n_\X,n^i_\X) \ .
\ee
Minimizing the action associated with ${\cal L}$ in the usual way will 
necessarily lead to the conservative Euler-Lagrange equations. \ To 
incorporate dissipation, we use Lagrange's generalized approach 
(which is championed by Carter, see for example
\cite{carter89:_covar_theor_conduc}), which is to obtain the 
Euler-Lagrange equations in the usual way, but instead of equating them 
to zero we put them equal to an appropriate sum of generalized forces
that may include dissipation. \ The net effect is that the Lagrangian 
serves to define the momenta, whose identification is perhaps the most 
crucial element of the multi-fluid construction that fully incorporates 
entrainment.

The first step towards identifying the unknown quantities is to write the 
total energy density ${\cal U}$ (to be distinguished from the total 
{\em internal} energy density ${\cal E}$) in terms of the fundamental 
variables of the action. \ Recall that a Hamiltonian density ${\cal H}$ 
can be generated as a Legendre transformation on ${\cal L}$, i.e.
\be
   {\cal H} \equiv \sum_\X n^i_\X p^\X_i - {\cal L} \ . \label{hamil}
\ee
The momenta are defined via
\be
   p^\X_i \equiv \left.\frac{\partial {\cal L}}{\partial n^i_\X}
   \right|_{n_\X} = g_{i j} m^\X v^j_\X - \left.\frac{\partial {\cal E}}
   {\partial n^i_\X}\right|_{n_\X} \ , \label{canmom}
\ee
where $v^i_\X = n^i_\X/n_\X$ is  the transport velocity of the 
$\X^{\rm th}$-constituent. \ The simplest, and most natural, way to 
obtain the total energy density in terms of the fundamental variables is 
to set ${\cal U} = {\cal H}$.

Accepting that, we now take a time-derivative of ${\cal H}$, and use 
Eqs.~(\ref{numx}) and (\ref{eulerx}), to find 
\bea
    \partial_t {\cal U} = \partial_t{\cal H} &=& - \nabla_i \left\{
                     \sum_\X \left(\left[\overline{\mu}{}^\X + 
                     \frac{1}{2} m^\X v^2_\X - v^j_\X p^\X_j\right] 
                     n^i_\X + v^j_\X T^\X{}^i{}_j\right)\right\}  \cr
                  && \cr
                  && + \sum_\X \left(v^i_\X \overline{f}{}^\X_i + 
                     T^\X{}^j{}_i \nabla_j v^i_\X + n^i_\X \nabla_i \left[
                     \overline{\mu}{}^\X + \frac{1}{2} m^\X v^2_\X - 
                     v^j_\X p^\X_j\right]  \right.\cr
                  && \cr
                  && \left. + \left[\overline{\mu}{}^\X + \frac{1}{2} 
                     m^\X v^2_\X - v^i_\X p^\X_i\right] \Gamma_\X\right) 
                     \ ,
\eea
where $v^2_\X = g_{i j} v^i_\X v^j_\X$ and
\be
   \overline{\mu}{}^\X \equiv \left.\frac{\partial {\cal E}}{\partial 
                         n_\X}\right|_{n^i_\X} \ .
\ee
Comparing with Eq.~(\ref{engcon}) we can identify
\bea
     Q^i &=& \sum_\X \left(\left[\overline{\mu}{}^\X + \frac{1}{2} m^\X 
             v^2_\X - v^j_\X p^\X_j\right] n^i_\X + v^j_\X T^\X{}^i{}_j
             \right) \ , \\
          && \cr
    \epsilon^{\rm ext} &=& \sum_\X \left(v^i_\X \overline{f}{}^\X_i + 
                       T^\X{}^j{}_i \nabla_j v^i_\X + n^i_\X \nabla_i 
                       \left[\overline{\mu}{}^\X + \frac{1}{2} m^\X 
                       v^2_\X - v^j_\X p^\X_j\right]  \right.\cr
                  && \cr
                  && \left. + \left[\overline{\mu}{}^\X + \frac{1}{2} 
                     m^\X v^2_\X - v^i_\X p^\X_i\right] \Gamma_\X\right) 
                     \ . \label{epsext}
\eea

It is useful at this point to note that there is a more complete Legendre 
transformation on ${\cal L}$ that can be made, by using the functions 
$p^\X_0$ defined as
\be
   p^\X_0 \equiv \left.\frac{\partial {\cal L}}{\partial n_\X}
   \right|_{n^i_\X} = - \left(\overline{\mu}{}^\X + \frac{1}{2} m^\X 
   v^2_\X\right) \ .
\ee
This leads to the new function $\Psi$ given by
\be
    \Psi \equiv - \left(\sum_\X \left[n_\X p^\X_0 + n^i_\X p^\X_i\right] 
                - {\cal L}\right) \ .
\ee
We will now establish that $\Psi$ enters the total momentum conservation 
equation as a generalized pressure. \ To do this we need to analyze more 
closely the stress-tensor $T^j{}_i$. 

Widely discussed in the general relativity literature (for instance, 
Ref.~\cite{mtw73}), is the fact that the stress-energy-momentum tensor 
can be obtained from a variation of the Lagrangian with respect to the 
metric. \  It is natural that metric variations are associated with 
stresses, since stress leads to deformations and the metric is what gives 
distances and volumes. \ In Appendix A we provide a non-rigorous account 
(using energy conservation arguments) of how to obtain the conservative 
piece of $T^j{}_i$, to be denoted $C^j{}_i$, by varying ${\cal L}$ with 
respect to $g_{i j}$ (see \cite{prix04:_multi_fluid} for a rigorous 
derivation). \ Our variation is to be understood as a displacement, in a 
small amount of time, of each fluid element within a fixed control 
volume, keeping the individual particle numbers and their associated 
velocities fixed. 

The discussion in Appendix~A prompts us to consider a less general form 
of the internal energy, i.e.
\be
    {\cal E} = {\cal E}\left(n_\X,w^2_{\X \Y}\right) \ , \label{eiso}
\ee
where
\be
    w^i_{\X \Y} = v^i_\X - v^i_\Y \quad , \quad w^2_{\X \Y} = g_{i j} 
                  w^i_{\X \Y} w^j_{\X \Y} \ .
\ee
This is manifestly Galilean invariant and locally isotropic. \ It is 
worth pointing out that this is not the most general form for the energy 
that we can allow. \ For example, it would be straightforward to include
an explicit dependence on the vorticity, eg.~in terms of the ``vorticity'' 
vector $W^i_{\X \s}$ (defined below by Eq.~(\ref{vort})) we could allow 
the energy to depend on $W_{\X \s}^2 = g_{i j} W^i_{\X \s} W^j_{\X \s}$. 
\ This would be analogous to the superfluid Helium model discussed by 
Bekarevich and Khalatnikov \cite{bekarevich61:_phenom_vortex} where the 
circulation around a vortex contributes to the energy budget, and would 
lead to terms that could be interpreted as the ``vortex tension''. \ 
Although one can think of other such possibilities we prefer to work with 
Eq.~(\ref{eiso}) here. \ Once the complete framework has been developed 
for this case, it should be relatively easy to extend it to include more 
general cases.

From Appendix A we then find that $C^j{}_i$ is given simply as 
\be
    C^j{}_i = \Psi \delta^j{}_i + \sum_\X n^j_\X p^\X_i \ . \label{consC}
\ee 
Thus, $\Psi$ behaves as the total pressure. \ We are naturally led to 
write the full stress-tensor as a sum of $C^j{}_i$ and a dissipation 
piece $D^j{}_i$, i.e.
\be
    T^j{}_i = C^j{}_i + D^j{}_i \ . \label{stressdecomp}
\ee
At this point it is important to note that, since the pressure $\Psi$ 
is not generally separable there is in general no way to write $T^j{}_i$ 
as just a sum of the $T^\X{}^j{}_i$. \ In other words, there is in 
general no useful notion of partial pressures, a point which is examined 
in more detail in Appendix B. \ Basically, it is more natural to work 
with the individual chemical potentials. \ Then the final equations 
retain the natural symmetry between different chemical constituents.  

The ramifications of a lack of partial pressures can be extracted by 
returning to Eq.~(\ref{consist}), and inserting Eq.~(\ref{stressdecomp}). 
\ We then get
\be
    f_i - \sum_\X f^\X_i = \nabla_j \left(D^j{}_i - \sum_\X \left[
    T^\X{}^j{}_i - n^j_\X p^\X_i\right]\right) \ ,
\ee
where we have defined 
\be
    \overline{f}{}^\X_i = f^\X_i + n_\X \nabla_i p^\X_0 + n^j_\X \nabla_i 
                          p^\X_j \ ,
\ee
and used the fact that
\be
    \nabla_i \Psi = - \sum_\X \left(n_\X \nabla_i p^\X_0 + n^j_\X 
    \nabla_i p^\X_j\right) \ .
\ee
Letting $T^\X{}^j{}_i$ be a sum of a conservative and dissipative piece, 
i.e.
\be
    T^\X{}^j{}_i = C^\X{}^j{}_i + D^\X{}^j{}_i \ , 
\ee
the consistency relation Eq.~(\ref{consist}) becomes
\be
    f_i - \sum_\X f^\X_i = \nabla_j \left(D^j{}_i - \sum_\X D^\X{}^j{}_i
    \right) - \sum_\X \nabla_j \left(C^\X{}^j{}_i - n^j_\X p^\X_i\right) 
    \ .
\ee

The key point is that we have now isolated terms in such a way that 
consistency of Eqs.~(\ref{momcon}) and (\ref{eulerx}) is guaranteed by 
taking
\be
     f_i = \sum_\X f^\X_i 
           \ , \quad
     D^j{}_i = \sum_\X D^\X{}^j{}_i 
           \ , \quad
     C^\X{}^j{}_i = n^j_\X p^\X_i \ . \label{bal}
\ee
Here we recall the discussion of Sec.~\ref{angmomcons} in which it was 
demonstrated that the total stress tensor $T^j{}_i$ is symmetric. \ Since 
the conservative piece $C^j{}_i$ can be shown to be symmetric, by using 
Eqs.~(\ref{canmom}) and (\ref{eiso}) in Eq.~(\ref{consC}), this means 
that we must require $D^j{}_i$ to be symmetric. \ However, as we already 
stated following Eq.~(\ref{ldd}), this does not mean that we should 
assume that the individual $D^\X{}^j{}_i$ are symmetric.
 
Finally, the equation of motion for the $\X^{\rm th}$-constituent takes 
the form 
\be
    \partial_t{\pi}{}^\X_i + \nabla_j \left(n^j_\X p^\X_i + D^\X{}^j{}_i
    \right) - \left(n_\X \nabla_i p^\X_0 + n^j_\X \nabla_i p^\X_j\right) 
    = f^\X_i \ , \label{eom1}
\ee
while the total external energy creation rate per unit volume becomes, 
cf. Eq.~(\ref{epsext}),
\be
    \epsilon^{\rm ext} = \sum_\X \left(v^i_\X f^\X_i + D^\X{}^j{}_i 
    \nabla_j v^i_\X - \left[p^\X_0 + v^i_\X p^\X_i\right] \Gamma_\X
    \right) \ . \label{gtot}
\ee

One can verify that our equations are consistent with those of earlier 
formulations, with the exception that our formulation has led to a 
balance of forces, Eq.~(\ref{bal}), that does not include the various 
dissipation stress-tensors $D^\X{}^j{}_i$. \ Thus, to complete the system 
we can specify each $D^\X{}^j{}_i$, and each $f^\X_i$ independently, with 
only the latter having to add to the total external force density $f_i$ 
acting on the whole system. \ We will now show how to limit further the 
possible forms for the $f^\X_i$ and $D^\X{}^j{}_i$, by employing 
Onsager's formulation \cite{onsager31:_symmetry} for multi-fluid systems.

\section{The Onsager Formulation for Dissipative Multi-Fluids} 
\label{onsager}

Because we have many independent fluids, the number of potential 
dissipation coefficients can be quite large. \ However, Onsager 
\cite{onsager31:_symmetry} demonstrated long ago that microscopic 
reversibility implies certain equalities among ``off-diagonal'' pieces of 
the entropy creation rate. \ Since this argument plays a key role in our 
analysis it is worth outlining the main ideas. \ A more detailed, 
pedagogical, description can be found in \cite{prigogine}. 

We begin by noting that the entropy, here represented by the number 
density $n_\s = s$, is maximal for a system in equilibrium. \ This means 
that any perturbation away from the equilibrium must be represented by 
quadratic deviations. \ Specifically, in the ``thermal frame'' associated 
with the entropy velocity $v_\s^i$, the conservation law Eq.~(\ref{numx}) 
implies
\be
    \frac{\Delta s}{\Delta t} \approx \frac{s - s_{\rm eq}}{\Delta t} 
                              \approx \Gamma_\s \ .
\ee
Comparing this to the anticipated expansion, see Jaynes 
\cite{jaynes85:_predict} for an elegant exposition of this, near an 
equilibrium
\be
   s \approx s_{\rm eq} - \frac{\Delta t}{2 T} \sum_{a,b} X_a L^{a b} X_b 
             \ , \label{sneq}
\ee 
we can identify \cite{hanley69:_transport}
\be
   T \Gamma_\s = - \frac{1}{2} \sum_{a,b} X_a L^{ab} X_b = \sum_{a = 1}^N 
                 J^a X_a \ , \label{ent_exp}
\ee  
where the $N$ individual $X_a$ are known as ``thermodynamic forces'' and 
the $J^a$ as ``fluxes.'' \ The thermodynamic forces represent a measure 
of the departure from global equilibrium in the system, with the fluxes 
arising in response. 
The Onsager symmetry principle simply states that $L^{a b} = L^{b a}$.  

Let us now see how we can apply this idea to our formalism. \ Our entropy 
creation rate is obtained from Eq.~(\ref{gtot}), by solving for 
$\Gamma_\s$. \ Noting that the chemical potentials $\mu^\X$, which are 
obtained from
\be
    \mu^\X \equiv \left.\frac{\partial {\cal E}}{\partial n_\X}
    \right|_{w^2_{\X \Y}} \ , 
\ee
are related to the $\overline{\mu}{}^\X$ via
\be
   \overline{\mu}{}^\X = \mu^\X - m^\X v^2_\X + v^i_\X p^\X_i \ ,
\ee
and that the temperature is given by $\mu^\s = T$, we can write
\be
    \epsilon^{\rm ext} = f_i v^i_\s + D^j{}_i \nabla_j v^i_\s + T 
                         \Gamma_\s + \sum_{\X \neq \s} \left(\Gamma_\X 
                         \mu^\X + \hat{f}{}^\X_i w^i_{\X \s} + 
                         D^\X{}_{i j} \nabla^i w^j_{\X \s}\right) 
                         \ . \label{tgammas}
\ee
We recall that $f_i$ and $D^j{}_i$ are the total force and dissipation 
tensors, respectively. We have assumed the locally isotropic, manifestly 
Galilean-invariant form for ${\cal E}$ (i.e.~Eq.~(\ref{eiso})), and we 
have defined
\be
    \hat{f}{}^\X_i \equiv f^\X_i - \frac{1}{2} m^\X \Gamma_\X g_{i j} 
                          \left(v^j_\X + v^j_\s\right) \ .
\ee
We will take the total system to be closed, which means that 
$\epsilon^{\rm ext} = 0$ and $f_i = 0$. \ In order to apply 
(\ref{ent_exp}) we also need to work in the entropy frame, which means 
all velocities will appear as the difference $w^i_{\X\s} = v^i_\X - 
v^i_\s$. 

The ``thermodynamic forces,'' which drive the system to equilibrium in 
the sense of Eq.~(\ref{sneq}), are seen from Eq.~(\ref{tgammas}) to be 
(for $\X \neq \s$)  
\be
   X_a = \left\{ 
         \begin{matrix}
              \{X^\X \equiv \mu^\X\} \ , \quad a = 1 \cr
              \{X_\X^i \equiv w^i_{\X \s}\} 
              \ , \quad a = 2 \cr   
              \{X_\X^{i j} \equiv \nabla^i w^j_{\X \s}\} 
              \ , \quad a = 3  
         \end{matrix}
         \right. \ ,
\ee
implying that the corresponding ``fluxes'' are (again for $\X \neq \s$)
\be
   J^a = \left\{ 
         \begin{matrix}
              \{J_\X \equiv - \Gamma_\X\} 
              \ , \quad a = 1 \cr
              \{J^\X_i \equiv - \hat{f}{}^\X_i\} 
              \ , \quad a = 2 \cr   
              \{J^\X_{i j} \equiv - D^\X{}_{i j}\} 
              \ , \quad a = 3  
         \end{matrix}
         \right. \ .
\ee
The Onsager formulation for our system is thus 
\bea
    J_\X &=& \sum_{\Y \neq \s} \left(L_{\X \Y} \mu^\Y + 
             \tilde{L}^{\X \Y}_i w^i_{\Y \s} + 
             \tilde{L}^{\X \Y}_{i j} \nabla^i w^j_{\Y \s}\right) \ , \cr
    J^\X_i &=& \sum_{\Y \neq \s} \left(\tilde{L}^{\X \Y}_i \mu^\Y + 
                 L^{\X \Y}_{i j} w^j_{\Y \s} + 
                 \tilde{L}^{\X \Y}_{i j k} \nabla^j w^k_{\Y \s}\right) 
                 \ , \cr
    J^\X_{i j} &=& \sum_{\Y \neq \s} \left(\tilde{L}^{\X \Y}_{i j} 
                     \mu^\Y + \tilde{L}^{\X \Y}_{i j k} w^k_{\Y \s} + 
                     L^{\X \Y}_{i j k l} \nabla^k w^l_{\Y \s}\right) \ .
\eea
The tensorial aspects of the ``$L$'' and ``$\tilde{L}$'' coefficients are 
handled by assuming that they can only be constructed from combinations 
of the thermodynamic forces, $\mu^\X$,  $w^i_{\X \s}$ and $\nabla^i 
w^j_{\X \s}$, and the background geometry terms, i.e.~the metric 
$g^{i j}$ and the volume form $\epsilon_{i j k} = \sqrt{g} [i j k]$, 
where $[i j k]$ is completely antisymmetric with $[1 2 3] = 1$. \ 
Moreover, in keeping with the spirit of the Onsager expansion and due to 
the fact that we are supposedly close to equilibrium, we consider only 
those terms that lead to quadratic combinations of the thermodynamic 
forces in $\Gamma_\s$. \ With these restrictions it is clear one would 
not expect coupling between forces with different tensorial nature. \ We 
readily find that $\tilde{L}^{\X \Y}_i =\tilde{L}^{\Y \X}_i = 0$ and 
that we can  rule out the use of $w^i_{\X \s}$ and $\nabla^i w^j_{\X \s}$ 
in any of the coefficients. \ The most general coefficients within our 
assumptions that can be written are
\bea
    L_{\X \Y} &=& \gamma_{\X \Y} = \gamma_{\Y \X} 
    \ , \cr
    \tilde{L}^{\X \Y}_{i j} &=& \tau^{\X \Y} g_{i j} = \tau^{\Y \X} 
    g_{i j} \ , \quad 
    L^{\X \Y}_{i j} = 2 {\cal R}^{\X \Y} g_{i j} = 2 {\cal R}^{\Y \X} 
    g_{i j} 
    \ , \cr
    \tilde{L}^{\X \Y}_{i j k} &=& {\cal A}^{\X \Y} \epsilon_{i j k} = 
    {\cal A}^{\Y \X} \epsilon_{i j k} \ , \cr 
    L^{\X \Y}_{i j k l} &=& \zeta^{\X \Y} g_{i j} g_{k l} +  
    \eta^{\X \Y} \left(g_{i k} g_{j l} +g_{i l} g_{j k} - \frac{2}{3} 
    g_{i j} g_{k l}\right) + \frac{1}{2} \sigma^{\X \Y} \epsilon_{i j m} 
    \epsilon^m{}_{k l} \cr
    &=& \zeta^{\Y \X} g_{i j} g_{k l} + \eta^{\Y \X} \left(
    g_{i k} g_{j l} + g_{i l} g_{j k} - \frac{2}{3} g_{i j} g_{k l}
    \right) + \frac{1}{2} \sigma^{\Y \X} \epsilon_{i j m} 
    \epsilon^m{}_{k l} \ .
\eea
Clearly, each coefficient inherits for its spatial indices the symmetries 
of either $g_{i j}$ or $\epsilon_{i j k}$. \ There has also been a lot of 
thought put into the coefficient $L^{\X \Y}_{i j k l}$. \ The key point 
is that the most general four-index object that can be constructed is a 
linear combination of the three terms $g_{i j} g_{k l}$, $g_{i k} 
g_{j l}$, and $g_{i l} g_{j k}$ (noting that $\epsilon_{i j m} 
\epsilon^m{}_{k l} = g_{i k} g_{j l} - g_{i l} g_{j k}$).

Our discussion will be greatly facilitated by introducing the trace-free 
shear $\Theta^{i j}_{\X \s}$, vorticity $\omega^{i j}_{\X \s}$, and 
expansion $\Theta_{\X \s}$: 
\bea
    \Theta^{i j}_{\X \s} &=& \frac{1}{2} \left(\nabla^i w^j_{\X \s} + 
                             \nabla^j w^i_{\X \s} - \frac{2}{3} g^{i j} 
                             \Theta_{\X \s}\right) = \Theta^{j i}_{\X \s} 
                             \ , \quad \Theta_{\X \s} = \nabla_i 
                             w^i_{\X \s} \ ,\cr
    \omega^{i j}_{\X \s} &=& \frac{1}{2} \left(\nabla^i w^j_{\X \s} - 
                             \nabla^j w^i_{\X \s}\right) = - 
                             \omega^{j i}_{\X \s} \ . \label{decomp}
\eea
We will also need the vorticity vector that is dual to the vorticity, 
i.e.
\be
   W^i_{\X \s} = \frac{1}{2!} \epsilon^i{}_{j k} \omega^{j k}_{\X \s} 
            \quad \Leftrightarrow \quad
   \omega^{i j}_{\X \s} = \epsilon^{i j}{}_k W^k_{\X \s} \ . \label{vort}
\ee
This allows us to utilize the well-known result that
\be
    \nabla^i w^j_{\X \s} = \Theta^{i j}_{\X \s} + \epsilon^{i j}{}_k 
                           W^k_{\X \s} + \frac{1}{3} g^{i j} 
                           \Theta_{\X \s} \ .
\ee
Here we should emphasize that the proper definition of vorticity, for 
which one recovers the Kelvin-Helmholtz conservation theorem, is in terms 
of the momentum \cite{prix04:_multi_fluid,mutualf}.\ In the absence of 
entrainment the quantity we use and the conserved vorticity are 
identical; but when entrainment is present, the difference is crucial. 
\ In terms of the analysis presented here, our introduction of the 
``vorticity'' is purely a matter of convenience, as it helps in 
separating purely symmetric (in the spatial indices) from antisymmetric 
objects. 

The fluxes are now of the form 
\bea
    \Gamma_\X &=& - \sum_{\Y \neq \s} \left(\gamma_{\X \Y} \mu^\Y + 
                  \tau^{\X \Y} \Theta_{\Y \s}\right) \ , \\
    \hat{f}^\X_i &=& - 2 g_{i j} \sum_{\Y \neq \s} \left({\cal R}^{\X \Y} 
                     w^j_{\Y \s} + {\cal A}^{\X \Y} W^j_{\Y \s}\right) 
                     \ , \\
    D^\X{}_{i j} &=& - \sum_{\Y \neq \s} \left(g_{ij} \left[\tau^{\X \Y} 
                     \mu^\Y + \zeta^{\X \Y} \Theta_{\Y \s}\right] + 2 
                     \eta^{\X \Y} g_{i k} g_{j l} \Theta_{\Y \s}^{k l} + 
                     \epsilon_{i j k} \left[{\cal A}^{\X \Y} w^k_{\Y \s} 
                     + \sigma^{\X \Y} W^k_{\Y \s}\right] \right) \ , 
                     \label{Dij}
\eea  
and we see that the entropy creation rate (in the thermal frame) is
\bea
    T \Gamma_\s &=& \sum_{\X,\Y \neq \s} \left(\gamma_{\X \Y} \mu^\X 
                    \mu^\Y + 2 \tau^{\X \Y} \mu^\X \Theta_{\Y \s} + 
                    \zeta^{\X \Y} \Theta_{\X \s} \Theta_{\Y \s} + 2 
                    \eta^{\X \Y} g_{i k} g_{j l} \Theta^{i j}_{\X \s} 
                    \Theta^{k l}_{\Y \s}  \right.\cr
                 && \left. + 2 g_{i j} \left[{\cal R}^{\X \Y} w^i_{\X \s} 
                    w^j_{\Y \s}+ 2 {\cal A}^{\X \Y} w^i_{\X \s} 
                    W^j_{\Y \s} + \sigma^{\X \Y} W^i_{\X \s} W^j_{\Y \s}
                    \right]\right) \ . \label{entrate1}
\eea
It is easy to show, given the explicit transformations in 
\cite{prix04:_multi_fluid} that the chemical potentials are Galilean 
invariant. \ Since all other terms in the right-hand side of 
Eq.~(\ref{entrate1}) depend only on velocity differences we see that the 
entire expression is invariant. \ This is, of course, as it should be. \
After all, the difference between reversible ($T \Gamma_\s=0$) and 
irreversible ($T\Gamma_\s>0$) processes cannot depend on the observer.

There are several different strategies to adopt in demonstrating the 
positive-definite nature of $\Gamma_\s$. \ Common to all is the simple 
notion of producing terms that look like $V^T M V$, where $M$ is a 
$d \times d$ symmetric matrix, and $V$ is a $d$-dimensional vector. \ 
At that point basic theorems from linear algebra 
(cf.~\cite{strang80:_lin_alg}) can be employed that will yield the 
constraints under which the matrix coefficients lead to positive-definite 
quantities. \ In our case, the most natural way to gather together terms 
is to introduce two new vectors (in this linear algebra sense):
\be
    V_1 = \left[\begin{matrix}
                \{\mu^\X\} \cr
                \{\Theta_{\X \s}\}
                \end{matrix}\right] 
          \ , \quad 
    V_2 = \left[\begin{matrix}
                \{w^i_{\X \s}\} \cr
                \{W^i_{\X \s}\}
                \end{matrix}\right] \ ,
\ee
whose dimension is $d = 2 (N - 1)$. \ Likewise, the matrices that define 
the quadratic terms are
\be
    M_1 = \left[\begin{matrix}
                \{\gamma_{\X \Y}\} & \{\tau^{\X \Y}\} \cr
                \{\tau^{\X \Y}\} & \{\zeta^{\X \Y}\}
                \end{matrix}\right] 
          \ , \quad 
    M_2 = \left[\begin{matrix}
                \{{\cal R}^{\X \Y}\} & \{{\cal A}^{\X \Y}\} \cr
                \{{\cal A}^{\X \Y}\} & \{\sigma^{\X \Y}\}
                \end{matrix}\right] \ .
\ee
The term containing $\Theta^{i j}_{\X \s} \Theta^{k l}_{\Y \s}$ is 
obviously already of the appropriate form. \ Hence, once the necessary 
constraints from linear algebra are obtained on the matrix coefficients, 
i.e.~the dissipation coefficients $\gamma^{\X \Y}$, $\zeta^{\X \Y}$, 
etc., $\Gamma_s \geq 0$ can be guaranteed in the entropy frame. 

Finally, we return to the inertial frame, in which the entropy velocity 
can appear independently of the other velocities. \ If we consider the 
system to  be closed, then the entropy creation rate just picks up the 
additional term proportional to $D^j{}_i$. \ As $D^\s{}_{i j}$ still 
remains undefined, we can just as well work with $D_{i j}$. \ Because 
$D_{i j}$ is contracted with $\nabla^i v^j_\s$, and it must be symmetric, 
the only proposal consistent with our main assumptions is
\be
   D_{i j} = - \left(\zeta^{\rm tot} \Theta_\s g_{i j} + 2 \eta^{\rm tot} 
             g_{i k} g_{j l} \Theta_\s^{k l}\right) \ , \label{entrate2}
\ee
where $\Theta_\s$ and $\Theta^{i j}_\s$ are defined as in 
Eq.~(\ref{decomp}) but replacing $w^i_{\X \s}$ with $v^i_\s$. \ Using 
the previous, linear algebra-based line of reasoning, we can determine, 
in principle, the constraints on the dissipation coefficients 
$\zeta^{\rm tot}$ and $\eta^{\rm tot}$ that will guarantee the positivity 
of $D_{i j} \nabla^i v^j_\s$, without spoiling the positive-definite 
nature of the other terms. 

\section{The Neutron, Proton, Electron and Entropy System} \label{npee}

\subsection{The problem of neutron star dissipation}

At zero temperature, the outer core of a neutron star consists of three 
interpenetrating fluids: superfluid neutrons, superconducting protons, 
and a highly degenerate gas of normal fluid electrons. \ Because of 
electromagnetic coupling the charged components lock together on a 
timescale that is much shorter than, eg., the timescale of stellar 
rotation or oscillation. \ Hence, the problem usually reduces to a 
two-fluid system. \ At a finite temperature, the situation is considerably 
more complex. \ Most obviously, we need to account also for the entropy. 
\ Provided that the system is far below the various superfluid transition 
temperatures, the entropy is associated with the electrons. \ However, 
there will also be contributions from excitations of the neutron and 
proton condensates. \ Accounting for these requires more thought. \ Based 
on the studies of the analogous problem for superfluid Helium there would 
seem to be (at least) two possible strategies. \ The difference between 
them is nicely described by Geurst \cite{geurst88b}.  

In the first approach, championed by Carter and his collaborators (see 
for example \cite{prix04:_multi_fluid}), each superfluid component is 
made up of the condensate and a massless gas of excitations. \ In our 
formalism this would amount to associating an entropy component to both 
the neutrons and the protons. \ It is then not difficult to show that the 
``normal fluid density'' associated with the the 
$\X^{\rm th}$-constituent is directly proportional to the entrainment 
coefficient $\alpha^{\X \s}$ (cf.~\cite{prix04:_multi_fluid}). \ In 
principle, both the neutrons and the protons will thus contribute to the 
normal fluid. \ This effect becomes particularly important near the 
superfluid transition temperatures. \ Far below the transition 
temperature, the temperature is essentially zero and one would expect 
both $\alpha^{\n \s}$ and $\alpha^{\p \s}$ to be effectively zero. \ 
Anyway, in order to account for a ``normal'' part of each superfluid we 
need to begin with a four-fluid system. \ Otherwise there would be no 
explicit distinction between the entrainment coefficients, eg.~between 
those that couple the fluids to the entropy, and those 
(i.e.~$\alpha^{\n \p}$, $\alpha^{\n \e}$, and $\alpha^{\p \e}$) that 
couple the species of particles to each other. \ As the electrons are not 
superconducting, we can assume that $v^i_\e = v^i_\s$ which means that we 
can set $\alpha^{\e \s} = 0$.  

The second possibility is perhaps closer to the orthodox approach to 
superfluids \cite{landau59:_fluid_mech,geurst88b}, in which one 
introduces an ad hoc separation of the mass density into a ``superfluid'' 
density and a ``normal'' fluid density. \ In our framework, this would 
involve dividing each density $n_\X$ into a piece associated with the 
condensate and a piece corresponding to the quasiparticle excitations. \ 
This philosophy was used in the recent calculation of entrainment 
parameters by Gusakov and Haensel \cite{gusakov}.

At the end of the day, the separation of a fluid constituent into two 
pieces is purely formal. \ One could not conceivably separate the 
superfluid condensate from the excitations in a real physical system. \ 
The mathematical formalism that we have developed should be flexible 
enough to allow us to represent both approaches to the finite temperature 
problem. \ As we will discuss elsewhere, the main issue concerns the 
\underline{interpretation} of the various variables (especially the 
entrainment coefficients).

It should also be noted that, despite neutron stars being 
self-gravitating bodies, we will not couple gravity to our model at this 
point. \ In the Newtonian context this is easily done by introducing 
an external force \cite{prix04:_multi_fluid}. \ We do not include this 
force here as our main focus is on the fluid aspects of the problem.

Let us proceed by assuming that there are four independent densities, 
eg.~the number densities for the neutrons, protons, and electrons, as 
well as the entropy density. \ These are denoted $(n_\n,n_\p,n_\e,s)$. \ 
A priori, the system will consist of four independent velocities, eg.~the 
neutron, proton, electron and entropy velocities which are denoted 
$(v^i_\n,v^i_\p,v^i_\e,v^i_\s)$. \ The entropy fluid is massless. \ We 
consider the four-fluid system first, in order to identify all the 
entrainment coefficients, before reducing to the case where all 
``normal'' fluids flow together. \ We believe that this strategy is more 
natural than one which imposes the existence of only two distinct 
transport velocities from the beginning.  

For the four-fluid system, the ``first law of thermodynamics'' is given 
by the expression
\bea
    {\rm d} {\cal E} &=& \mu_\n {\rm d} n_\n + \mu_\p {\rm d} n_\p + 
    \mu_\e {\rm d} n_\e + T {\rm d} s + \alpha^{\n \p} {\rm d} 
    w^2_{\n \p} + \alpha^{\n \e} {\rm d} w^2_{\n \e}  \cr
    &+& \alpha^{\e \p} {\rm d} w^2_{\e \p} + \alpha^{\n \s} {\rm d} 
    w^2_{\n \s} + \alpha^{\p \s} {\rm d} w^2_{\p \s} + \alpha^{\e \s} 
    {\rm d} w^2_{\e \s} \ , \label{firstlaw}
\eea
where we are again assuming the Galilean-invariant, locally isotropic 
form for ${\cal E}$. \ Because of overall charge neutrality we have
\be
   n_\p = n_\e \ .
\ee
This, in conjunction with Eqs.~(\ref{numx}) and (\ref{pcreacon}), implies
\be
    \Gamma_\p = \Gamma_\e = - \Gamma_\n 
\ee
and
\be
    m^\n = m^\p + m^\e \equiv m \ ,
\ee 
where $m^\n$, $m^\p$, and $m^\e$ are the neutron, proton and electron 
masses, respectively. \ The electrons should also closely track the 
protons because of the electromagnetic attraction 
\cite{alpar84:_rapid_postglitch,mendell91:_nondiss}. \ Thus, it is a good 
approximation to simply set
\be
   v^i_\p = v^i_\e \equiv v^i_\c \ .
\ee
In a later work, we will consider electromagnetic coupling in more detail 
in order to justify  fully this constraint (see also 
\cite{mendell91:_coupl_charg}).

In the absence of thermal conduction $v^i_\s = v^i_\c$. \ Because of this 
the entropy flux is often written as
\be
   s v^i_\s = s v^i_\c + \frac{1}{T} q^i \implies q^i = s T w^i_{\s \c} 
              \ ,
\ee
where $q^i$ is the heat-flux vector. \ However, by introducing the 
heat-flux vector we break the natural symmetry of our formalism since 
$v^i_\c$ is put on a special footing. \ In general $q^i$ is dynamically 
independent of the other two velocities. \ A standard result is to argue 
that any deviations of $v^i_\s$ from co-motion with $v^i_\c$ will be due 
to gradients of the temperature. \ Because of the definition of 
temperature, cf.~Eq.~(\ref{firstlaw}), the entropy flux still 
represents a dynamically independent variable, since $T$ depends on the 
relative velocities $w^i_{\X \Y}$, which include $v^i_\s$. \ This kind of 
velocity dependence is manifested in the momenta $p^\X_i$, for instance, 
via the entrainment coefficients $\alpha^{\X \Y}$. 

It is convenient to define 
\be
    \alpha^{\n \c} \equiv \alpha^{\n \p} + \alpha^{\n \e} \ , \quad
    \alpha^{\c \s} \equiv \alpha^{\p \s} + \alpha^{\e \s} \ ,
\ee
and  
\be
   \mu^\c = \mu^\p + \mu^\e \ .
\ee
Now the three independent fluid momentum densities can be written
\bea
    \pi^\n_i &=& g_{i j} \left(m n_\n v^j_\n - 2 \left[\left(
                 \alpha^{\n \c} + \alpha^{\n \s}\right) w^j_{\n \c} - 
                 \alpha^{\n \s} w^j_{\s \c}\right]\right) \ ,  \\
         && \cr
    \pi^\c_i &\equiv& n_\p p^\p_i + n_\e p^\e_i = g_{i j} \left(m n_\c 
                      v^j_\c + 2 \left[\alpha^{\n \c} w^j_{\n \c} + 
                      \alpha^{\c \s} w^j_{\s \c}\right]\right) \ ,  \\
         && \cr
    \pi^\s_i &\equiv& 2 g_{i j} \left(\alpha^{\n \s} w^j_{\n \c} - 
                      \left[\alpha^{\n \s} + \alpha^{\c \s}\right] 
                      w^j_{\s \c}\right)\ .  
\eea
Defining
\be
   D^\c{}^j{}_i = D^\p{}^j{}_i + D^\e{}^j{}_i \ , 
\ee
we find for the three force densities, cf. (\ref{eom1}), 
\bea
     f^\n_i &=& \partial_t{\pi}{}^\n_i + \nabla_j \left(v^j_\n \pi^\n_i + 
                D^\n{}^j{}_i\right) + n_\n \nabla_i \left(\mu_\n - 
                \frac{1}{2} m v^2_\n\right) + \pi^\n_j \nabla_i v^j_\n 
                \ , \label{eulern} \\
        && \cr
     f^\c_i &=& \partial_t{\pi}{}^\c_i + \nabla_j \left(v^j_\c \pi^\c_i + 
                D^\c{}^j{}_i\right) + n_\c \nabla_i \left(\mu_\c - 
                \frac{1}{2} m v^2_\c\right) + \pi^\c_j \nabla_i v^j_\c 
                \ , \label{eulerc} \\
        && \cr
     f^\s_i &=& \partial_t{\pi}{}^\s_i + \nabla_j \left(v^j_\s \pi^\s_i + 
                D^\s{}^j{}_i\right) + s \nabla_i T + \pi^\s_j \nabla_i 
                v^j_\s \ . \label{eulerq} 
\eea

We can put the finishing touches on our ``neutron star model'', by 
building the thermodynamic ``forces'' and ``fluxes.'' \ Key to this will 
be the assertion that the thermodynamic forces are to be considered as 
linearly independent, i.e.~if a linear combination of thermodynamic 
forces is to be equal to zero, then the coefficients multiplying each 
force must vanish. This assumption does not seem unreasonable.
\ We must first consider the constraints that result 
from the equalities among the particle creation rates $\Gamma_\n$, 
$\Gamma_\p$, and $\Gamma_\e$. 

By setting $\Gamma_\e = \Gamma_\p = - \Gamma_\n$, and imposing linear 
independence, we determine that 
\be
    \gamma_{\n \n} = - \gamma_{\p \n} = - \gamma_{\e \n} = \gamma_{\p \p} 
                   = \gamma_{\e \p} = \gamma_{\e \e}
\ee
and
\begin{eqnarray}
    \tau^{\n \n} &=& - \tau^{\n \p} = - \tau^{\n \e} \ , \\
    2  \tau^{\n \n} &=&  \tau^{\p \p} + \tau^{\p \e} \ , \\
    \tau^{\p \p} &=& \tau^{\e \e} \ .
\end{eqnarray}

If we define
\be
    \begin{array}{ll} {\cal X}^{\n \c} \equiv {\cal X}^{\n \p} + 
       {\cal X}^{\n \e} \ , \\ 
       {\cal X}^{\c \c} \equiv {\cal X}^{\p \p} + 2 
       {\cal X}^{\p \e} + {\cal X}^{\e \e} \end{array} 
       \ , \quad \mbox{ where } {\cal X} = {\cal R}\ , {\cal A}\ ,
       \zeta\ , \eta\ \mbox{ or }\sigma 
\ee
then the fluid forces are
\bea
     \hat{f}^\n_i &=& - 2 g_{i j} \left({\cal R}^{\n \n} w^j_{\n \s} + 
                      {\cal R}^{\n \c} w^j_{\c \s} +  {\cal A}^{\n \n} 
                      W^j_{\n \s} + {\cal A}^{\n \c} W^j_{\c \s}\right) 
                      \ , \\
     \hat{f}^\c_i &=& - 2 g_{i j} \left({\cal R}^{\c \n} w^j_{\n \s} + 
                      {\cal R}^{\c \c} w^j_{\c \s} + {\cal A}^{\c \n} 
                      W^j_{\n \s} + {\cal A}^{\c \c} W^j_{\c \s}\right) 
                      \ , 
\eea
and using the fact that we are considering a closed system, we also have
\be
     \hat{f}^\s_i = f^\s_i = - \left(\hat{f}^\n_i + \hat{f}^\c_i + 
                      \frac{1}{2} m \Gamma_\n g_{i j} w^j_{\n \c}
                      \right) \ .
\ee

Finally, the independent dissipation coefficients become, 
cf.~Eq.~(\ref{Dij}), 
\bea
     D^\n{}_{i j} &=& - \left(g_{i j} \left[\tau^{\n \n} \left(\mu^\n - 
                    \mu^\c\right) + \zeta^{\n \n} \Theta_{\n \s} + 
                    \zeta^{\n \c} \Theta_{\c \s}\right] + 2 g_{i k} 
                    g_{j l} \left[\eta^{\n \n} \Theta^{k l}_{\n \s}  +
                    \eta^{\n \c} \Theta^{k l}_{\c \s}\right] \right. \cr
                 && \left.  +\epsilon_{i j k} \left[{\cal A}^{\n \n} 
                    w^k_{\n \s} + {\cal A}^{\n \c} w^k_{\c \s} + 
                    \sigma^{\n \n} W^k_{\n \s} + \sigma^{\n \c} 
                    W^k_{\c \s}\right]\right) \ , \\
     D^\c{}_{i j} &=& - \left(g_{i j} \left[2 \tau^{\n \n} \left(\mu^\c - 
                    \mu^\n\right) + \zeta^{\c \n} \Theta_{\n \s} + 
                    \zeta^{\c \c} \Theta_{\c \s}\right] + 2 g_{i k} 
                    g_{j l} \left[\eta^{\c \n} \Theta^{k l}_{\n \s} + 
                    \eta^{\c \c} \Theta^{k l}_{\c \s}\right]  \right. \cr
                 && \left. + \epsilon_{i j k} \left[{\cal A}^{\c \n} 
                    w^k_{\n \s} + {\cal A}^{\c \c} w^k_{\c \s} + 
                    \sigma^{\c \n} W^k_{\n \s} + \sigma^{\c \c} 
                    W^k_{\c \s}\right]\right) \ , 
\eea
We also get
\bea
     D^\s{}_{i j} &=& D_{ij} - D^\n{}_{ij} - D^\c{}_{ij} =  - \left(
                      g_{i j} \left[\zeta^{\rm tot} \Theta_s - 
                      \tau^{\n \n} \left(\mu^\n - \mu^\c\right) - \left(
                      \zeta^{\n \n} + \zeta^{\c \n}\right) \Theta_{\n \s} 
                      - \left(\zeta^{\n \c} + \zeta^{\c \c}\right) 
                      \Theta_{\c \s}\right] \right. \cr 
                   &&  + 2 \left. g_{i k} g_{j l} 
                      \left[\eta^{\rm tot} \Theta^{k l}_\s - 
                      \left(\eta^{\n \n} + \eta^{\c \n}
                      \right) \Theta^{k l}_{\n \s} - \left(\eta^{\c \c} + 
                      \eta^{\n \c}\right) \Theta^{k l}_{\c \s}\right]  
                      \right. \cr 
                   && \left. - \epsilon_{i j k} \left[ \left(
                      {\cal A}^{\n \n} + {\cal A}^{\c \n}\right) 
                      w^k_{\n \s} + \left({\cal A}^{\c \c} + 
                      {\cal A}^{\n \c}\right) w^k_{\c \s} + 
                      \left(\sigma^{\n \n} + \sigma^{\c \n}\right) 
                      W^k_{\n \s} + \left(\sigma^{\c \c} +\sigma^{\c \n} 
                      \right) W^k_{\c \s}\right]\right) \ .
\eea
It is worth noting that this construction ensures that $D_{ij}$ is 
symmetric, as required for a closed system.

The final step is to use the positivity of the entropy to set constraints 
on the dissipation coefficients. \ It follows from Eqs.~(\ref{entrate1}) 
and (\ref{entrate2}), and some algebra, that the entropy creation rate is
\bea
    T \Gamma_\s &=& \zeta^{\rm tot} \Theta_\s^2 + 2 \eta^{\rm tot} 
                    g_{i k} g_{j l} \Theta^{i j}_\s \Theta^{k l}_\s \cr
                 && + 2 g_{i k} g_{j l} \left[\begin{matrix}
                    \Theta^{i j}_{\n \s} &  \Theta^{i j}_{\c \s}
                    \end{matrix}\right] \left[\begin{matrix}
                    \eta^{\n \n} & \eta^{\n \c} \cr  
                    \eta^{\n \c} & \eta^{\c \c}
                    \end{matrix}\right] \left[\begin{matrix}
                    \Theta^{k l}_{\n \s} \cr  \Theta^{k l}_{\c \s}
                    \end{matrix}\right]  \cr
                 && \cr 
                 && + \left[\begin{matrix}
                    \mu^\c - \mu^\n &  \Theta_{\n \s} & \Theta_{\c \s}
                    \end{matrix}\right] \left[\begin{matrix}
                    \gamma_{\n \n} & - \tau^{\n \n} & 2 \tau^{\n \n} \cr  
                    - \tau^{\n \n} & \zeta^{\n \n} & \zeta^{\n \c} \cr
                    2 \tau^{\n \n} & \zeta^{\n \c} & \zeta^{\c \c}
                    \end{matrix}\right] \left[\begin{matrix}
                    \mu^\c - \mu^\n \cr \Theta_{\n \s} \cr  \Theta_{\c \s}
                    \end{matrix}\right] \cr 
                 && \cr
                 && + 2 g_{i j} \left[\begin{matrix}
                     w^i_{\n \s} & w^i_{\c \s} & W^i_{\n \s} & W^i_{\c \s}
                    \end{matrix}\right] \left[\begin{matrix}
                    {\cal R}^{\n \n} & {\cal R}^{\n \c} & 
                    {\cal A}^{\n \n} & {\cal A}^{\n \c} \cr  
                    {\cal R}^{\n \c} & {\cal R}^{\c \c} & 
                    {\cal A}^{\n \c} & {\cal A}^{\c \c} \cr
                    {\cal A}^{\n \n} & {\cal A}^{\n \c} & \sigma^{\n \n} 
                    & \sigma^{\n \c} \cr
                    {\cal A}^{\n \c} & {\cal A}^{\c \c} & \sigma^{\n \c} 
                    & \sigma^{\c \c}\end{matrix}\right] 
                    \left[\begin{matrix}
                    w^j_{\n \s} \cr w^j_{\c \s} \cr W^j_{\n \s} \cr 
                    W^j_{\c \s} \end{matrix}\right]\ .
\eea
This expression shows that the problem is extremely rich. \ Our system 
has in general $19$ independent dissipation coefficients. \ This can be 
compared to the general case for superfluid Helium which, according to 
Putterman \cite{putterman74:_superfluid}, requires 13 independent 
coefficients.

In principle, it is now a straightforward matter to determine the 
constraints on the dissipation coefficients, since each defining matrix 
is symmetric and real, and thus linear-algebra theorems can be employed 
to determine when each contribution is positive definite, specifically, 
if all the eigenvalues of the defining matrix are positive then the 
quadratic form will be positive definite. \ For example, it is not 
difficult to show that if both $\eta^{\n \n}$ and $\eta^{\c \c}$ are 
negative then there are no values for the coefficients that will lead to 
a positive-definite form. \ On the other hand, if $\eta^{\n \n} \geq 0$ 
and $\eta^{\c \c} \geq 0$, then the quadratic form is positive if and 
only if
\be
    \eta^{\n \n} \eta^{\c \c} > \left(\eta^{\n \c}\right)^2 \ .
\ee
A similar analysis for the second and third matrices would be tractable 
analytically, since the characteristic equations for the eigenvalues will 
be third and fourth order, respectively. \ However, it is not clear that 
working out the detailed conditions would add any further insight at the 
present time. \ After all, many of the coefficients in our general 
expression are new, and it may be more productive to begin by seeing if 
we can interpret their physical meaning.   

\subsection{Final reduction: The two-fluid equations}

In order to connect with previous work on superfluid neutron stars we 
conclude our discussion by making one further reduction. \ We will 
neglect heat conduction and let the entropy flow with the charged fluid 
components. \ This simply means that we take $v_\s^i = v_\c^i$ in all 
equations in the previous section. \ This implies that we only have two 
independent velocities in the problem, and hence only need two equations 
of motion. \ Given this, it is natural to combine the momenta of the 
entropy and the charged components and consider
\be
    \tilde{\pi}_i^\c = \pi_i^\c + \pi_i^\s = g_{i j} \left[m n_\c v_\c^j 
    + 2 \left(\alpha^{\n \c} + \alpha^{\n \s}\right) w_{\n \c}^j\right]
    = m n_\c g_{ij} \left(v_\c^j + \varepsilon_\c w_{\n \c}^j\right)
\ee
where we have defined the entrainment parameter
\be
    \varepsilon_\c = \frac{2 \left(\alpha^{\n \c} + \alpha^{\n \s}\right)}
                     {m n_\c} \ .
\ee

Next, combining Eqs.~(\ref{eulerc}) and (\ref{eulerq}) we see that
\bea
     \partial_t{\tilde{\pi}}_i^\c &+& \nabla_j \left(v_\c^j 
     \tilde{\pi}_i^\c\right) + \tilde{\pi}_j \nabla_i v_\c^j + n_\c 
     \nabla_i \left(\mu_\c - \frac{1}{2} m v_\c^2\right) + s \nabla_i T 
     \cr
     &=& - \hat{f}^\n_i - \frac{1}{2} m \Gamma_\n w_i^{\n \c} - m 
     \Gamma_\n v_i^\c - \nabla_j\left(D^\c{}^j{}_i + D^\s{}^j{}_i\right) 
     \ .
\eea
Rewriting this equation in terms of the explicit transport velocities, 
as in \cite{prix04:_multi_fluid,andersson04:_canon_energy}, we obtain
\bea
     && m n_\c \left(\partial_t + \pounds_{v_\c}\right) \left[v^\c_i + 
        \varepsilon_\c w^{\n \c}_i\right] + n_\c \nabla_i \left(\mu_\c - 
        \frac{1}{2} m v_\c^2\right) + s \nabla_i T \cr
     && \qquad = - \hat{f}^\n_i - \frac{1}{2} m \left(1 - 2 
        \varepsilon_\c\right) \Gamma_\n w_i^{\n \c} + \nabla_j \left(
        D^\n{}^j{}_i - D^j{}_i\right) \ , \label{eqc}
\eea
where we have used the fact that the Lie derivative (which measures the 
rate at which a quantity changes relative to its motion) of a covector 
$a_i$ follows from
\be
    \pounds_v a_i = v^j \nabla_j a_i + a_j \nabla_i v^j \ .
\ee
We have also made use of Eq.~(\ref{numx}).

The corresponding equation for the neutrons follows from 
Eq.~(\ref{eulern}). \ We get
\bea
     && m n_\n \left(\partial_t + \pounds_{v_\n}\right) \left[v^\n_i + 
        \varepsilon_\n w^{\c \n}_i\right] + n_\n \nabla_i \left(\mu_\n - 
        \frac{1}{2} m v_\n^2\right) \cr
     && \qquad = \hat{f}^\n_i + \frac{1}{2} m \left(1 - 2 
        \varepsilon_\n\right) \Gamma_\n w_i^{\c \n} - \nabla_j 
        D^\n{}^j{}_i \ , \label{eqn}
\eea
where 
\be
    \varepsilon_\n = \frac{ 2 \left(\alpha^{\n \c} + \alpha^{\n \s}
                     \right)}{m n_\n} \ .
\ee

The two equations of motion, Eqs.~(\ref{eqc}) and (\ref{eqn}), can be 
directly compared to  Eqs.~(176) and (177) in \cite{prix04:_multi_fluid}. 
\ If we introduce 
\be
    f_i^{\rm mut} = \hat{f}^\n_i + \frac{1}{2} m \left(1 - 2 
                    \varepsilon_\n\right) \Gamma_\n w_i^{\c \n}
\ee
we find that the two results agree perfectly in the zero-temperature 
($T = 0$), non-dissipative ($D^\n{}_{ij} = D_{ij}=0$) limit. \ Thus our 
representation reduces to the familiar result in the relevant limit. \ In 
the more general case that we have considered, several new aspects appear 
naturally. \ Finite temperature effects are explicitly represented via 
the presence of the $s \nabla_i T$ term in the equation for the charged 
fluid/entropy equation. \ Implicitly, we are also accounting for the 
possibility of a normal part of the neutrons via the entrainment 
coefficient $\alpha^{\n \s}$. \ The two fluids are coupled in a number of 
ways. \ In particular, the mutual coupling via $\hat{f}_i^\n$ and 
$D^\n{}_{ij}$, which accords with Newton's Third Law, should be noted. \ 
The presence of the ``total'' dissipation in terms of $D_{ij}$ is also 
relevant. 

It is instructive to consider the above equations of motion in the 
context of what we already know about dissipative fluids. \ In 
particular, we would like to understand which of our many coefficients 
can be considered as known, and which represent potentially new aspects 
of the problem. \ To carry out this exercise, let us begin on familiar 
territory and consider the total dissipation part. \ From our 
definitions, we have
\be
    - \nabla^j D_{ij} = \nabla^j \left(\zeta^{\rm tot} \Theta_\s g_{i j} 
                        + 2 \eta^{\rm tot} g_{i k} g_{j l} 
                        \Theta_\s^{k l}\right) \label{Dij2}
\ee
which should be compared to the standard right-hand side of the 
Navier-Stokes equations \cite{landau59:_fluid_mech}
\be
    \nabla^j \left[\eta \left(\nabla_j v_i + \nabla_i v_i - \frac{2}{3} 
    g_{ij} \nabla_l v^l\right)\right] + \nabla_i \left(\zeta \nabla_l v^l
    \right)  = \nabla^j \left(2 \eta \Theta_{ij}\right) + g_{ij} \nabla^j 
    \left(\zeta \Theta\right) \ . 
\ee
Comparing the two expressions we can readily identify the coefficients of 
the standard shear and bulk viscosities as
\be
     \eta = \eta^{\rm tot} \qquad \mbox{ and } \qquad 
    \zeta = \zeta^{\rm tot} \ .
\ee

Not surprisingly, the interior coupling terms are not as straightforward 
to explain. \ This is quite obvious since we have tried to write down a 
more or less general expression for the permissible terms in a 
multi-fluid context where previous studies have been limited. \ The 
problem is exacerbated by the fact that the bulk of the extant work was 
carried out in the ``orthodox'' framework 
\cite{landau59:_fluid_mech}. \ To interpret the 
result, we consider the combination
\bea
    \hat{f}^\n_i - \nabla_j D^\n{}^j{}_i &=& - 2 {\cal R}^{\n \n} 
       w_i^{\n \c} + \epsilon_{i j k} \left(\nabla^j {\cal A}^{\n \n}
       \right) w_{\n \c}^k \cr
    && + \nabla^j \left\{ g_{i j} \left[\tau^{\n \n} \left(\mu_\n - \mu_\c
       \right) - \zeta^{\n \n} \Theta_{\n \c}\right] + 2 \eta^{\n \n} 
       \Theta_{i j}^{\n\c} + \sigma^{\n \n} \omega_{i j}^{\n \c}\right\} 
       \ , \label{interns}
\eea
which follows from the various definitions and some straightforward 
algebra. \ Given this expression we can interpret the individual terms in 
the following way: Since it is proportional to the relative velocity, the 
term involving ${\cal R}^{\n \n}$ is associated with a force analogous to 
standard resistivity. \ Meanwhile, ${\cal A}^{\n \n}$ relates to a
Magnus-type force which acts orthogonally to the relative flow 
\cite{mutualf}. \ The two terms involving $\Theta_{\n \c}$ and 
$\Theta_{i j}^{\n \c}$ are simply analogues of the standard bulk- and 
shear-viscosities, although here it is the expansion and shear of the 
relative flow which is important. \ The coefficient $\tau^{\n \n}$  
accounts for effects due to the fluids being driven out of chemical 
equilibrium by the flow. \ In recent work Carter and collaborators have 
refered to this type of interaction as ``transfusion''. \ The final term 
in Eq.~(\ref{interns}), related to $\sigma^{\n \n}$, appears to be new. 

To conclude our discussion, let us consider the particular form of the 
mutual friction force which arises because of a balance between i) the 
Magnus force acting on quantised superfluid neutron vortices and ii) 
scattering of electrons off of the vortices. \ As we have shown 
elsewhere \cite{mutualf}, this results in a force 
\be
    f^{\rm mf}_i = {\cal B}^\prime \rho_\n n_v \epsilon_{i j k} \kappa^j 
                   w_{\c \n}^k + {\cal B} \rho_\n n_v \epsilon_{i j k} 
                   \epsilon^{k l m} \hat{\kappa}^j \kappa_l w_m^{\c \n} 
                   \ , \label{fmf} 
\ee
acting on the neutrons. \ Here $n_v$ is the vortex surface density and 
$\kappa^i$ is a vector which is aligned with the vortices and has 
magnitude $h/2m_\n$ (a hat denotes a unit vector); see \cite{mutualf} for 
further details. 

Comparing this expression to Eq.~(\ref{interns}) we see that only the 
first two terms in the latter are relevant. \ In order to incorporate 
Eq.~(\ref{fmf}) in our framework, we need to be able to identify the 
coefficients ${\cal R}^{\n \n}$ and $\nabla_i {\cal A}^{\n \n}$. \ We 
easily see that we must have 
\be
    \nabla_i {\cal A}^{\n \n} = - {\cal B}^\prime \rho_\n n_v \kappa_i 
    \label{Bexp} \ .
\ee
The corresponding expression for ${\cal R}^{\n\n}$ follows from 
\be
   {\cal B} \rho_\n n_v \epsilon_{i j k} \epsilon^{k l m} \hat{\kappa}^j 
   \kappa_l w_m^{ \c\n} = {\cal B} \rho_\n n_v \kappa \left(\hat{\kappa}_i 
   \hat{\kappa}_j - g_{i j}\right) w_{\c\n}^j \  = 2 {\cal R}^{\n\n} 
   w_i^{\c\n} \ .
\ee
Multiplying each side by $w^i_{\c\n}$ we find that
\be
    {\cal R}^{\n\n} = \frac{1}{2}\rho_\n n_v \kappa {\cal B} \left[
                      (\hat{w}^i_{\c\n} \hat{\kappa}_i)^2 - 1\right] \ .
\ee
The above expressions can now be used, given the explicit estimates for 
the two coefficients ${\cal B}$ and ${\cal B}'$ obtained in 
\cite{mutualf}, to incorporate mutual friction in our dissipative neutron 
star model. \ This will allow us to extend our previous studies of 
superfluid neutron star oscillations and instabilities in a number of 
exciting directions.  

\section{Concluding Remarks}

We have described a general formalism which can be used to model 
dissipative multi-fluid systems. \ Our discussion was based on the key 
conservation laws for mass, energy and linear and angular momentum, 
together with recent progress in deriving the equations of motion for the 
non-dissipative problem from a variational principle 
\cite{prix04:_multi_fluid,carter04:_cov1,carter03:_cov2,carter04:_cov3}, 
and the classic Onsager symmetry principle which provides a strategy for 
including dissipative terms in such a way that the second law of 
thermodynamics is satisfied. \ Following this approach, we derived a set 
of multi-fluid equations which has a number of advantages over the 
``standard'' formalism used to model, in particular, superfluid He$^4$. \ 
The most important difference is that we have made due distinction 
between velocities and momenta, which means that the entrainment effect 
is accounted for in a natural way. 

As an application of the general formalism, we  considered two models for 
a neutron star core. \ The first model accounts for three distinct 
fluids, the superfluid neutrons, a conglomerate of charged particles 
(protons and electrons) and the entropy. \ We have shown that this model, 
in principle, requires 19 distinct dissipation coefficients to be 
determined. \ Detailed microscopic analysis is needed to understand the 
relevance of the majority of these coefficients. \ We also discussed how 
the normal fluid fraction of neutrons that should be present at finite 
temperatures can be associated with the entrainment parameter 
$\alpha^{\n \s}$ which encodes how the equation of state depends on the 
relative motion between the neutrons and the entropy. \ Our second model 
is obtained by neglecting heat conductivity. \ This couples the entropy 
to the charged fluid and reduces the number of independent degrees of 
freedom to two. \ We discussed the form of the various dissipative terms 
in this two-fluid model, and interpreted their physical meaning. \ We 
also showed how the mutual friction, which is mediated by the superfluid 
vortices, can be accounted for within our model.

One can envisage a number of interesting applications of the framework
that we have developed in this paper. \ In the case of neutron stars, the 
possibilities range from the viscous damping of neutron star oscillations 
due to electron-electron scattering \cite{andersson04:_viscosity_coeff} 
to intricate issues concerning the effect of entrainment on the mutual 
friction force \cite{mutualf}. \ Since our model allows for much more 
complicated multi-fluid dynamics than has so far been considered in the 
literature, one would have to be somewhat suspicious about any statements 
about the damping of superfluid neutron star oscillations which were made 
without consideration of the various degrees of freedom. \ It would 
certainly be interesting to revisit the problem of oscillation modes 
driven unstable by gravitational radiation \cite{Andersson:2002ch} and 
investigate the role of the true multi-fluid degrees of freedom. \ Our 
hope is this work will inspire a significant improvement of our 
understanding of superfluid neutron star dynamics, and perhaps generate 
input from other communities (eg.~chemistry) that have much experience 
with multi-fluid systems. 

Finally, it is worth emphasising that our formalism is quite general, and
one would expect it to be useful also in other problem areas. \ Applying 
it to other multi-phase flow problems, as discussed in for example
\cite{drew,geurst86,geurst88a}, seems relatively straightforward. \ There 
are also more exotic possibilities. In particular, we believe that our 
formalism could be quite relevant for problems involving fluid analogues 
of various curved spacetimes, such as black holes and big-bang 
cosmologies, that can be potentially observable in laboratory systems 
(see \cite{Comer:1991ci} for a discussion of ``sonic'' horizons in 
superfluids, and \cite{barcelo05:_livrev} for a recent review). \ For 
this kind of application, it would be particularly important to establish 
whether the  ``spacetime'' effects are likely to be rendered 
experimentally undetectable by various forms of dissipation.  

\acknowledgments

NA acknowledges support from PPARC via grant no PPA/G/S/2002/00038 and 
Senior Research Fellowship no.~PP/C505791/1. \ GLC gratefully 
acknowledges partial support from NSF grant PHYS-0457072. \ We are much 
indebted to Brandon Carter for the many discussions over the years on the 
subtleties of multi-fluid dynamics.

\section*{Appendix A}

Here we will motivate why the conservative part of the total stress 
tensor $C^j{}_i$ can be obtained from a variation of the action with 
respect to the metric. \ The analysis is by no means rigorous, but it 
does lead to the result that has been obtained rigorously in 
Ref.~\cite{prix04:_multi_fluid}. \ Our aim is to give the reader not 
familiar with all aspects of action-based field theory some basic 
physical intuition, using simple energy conservation arguments.

Let us consider a displacement $\delta \xi^i$, in a small amount of time 
$\delta t = t_2 - t_1$ of a closed, isolated system such that the 
individual constituent particle numbers  as well as their corresponding 
velocities are kept fixed. \ We will also suppose that each constituent 
undergoes the same displacement so that their respective velocities can 
be written
\be
    v^i_\X \approx \frac{\delta \xi^i}{\delta t} \ .
\ee
Because the system is closed and isolated, the total energy will be 
conserved. \ From Eqs.~(\ref{toteng}) and (\ref{hamil}) the change in 
total energy can be written
\bea
    \delta U &=& U(t_2) - U(t_1) \cr
              && \cr
             &=& \int_{t_1}^{t_2} {\rm d} t~\frac{{\rm d}}{{\rm d} t} 
                 \int_V \left(\sum_\X n^i_\X p^\X_i - {\cal L}
                 \right)~{\rm d} V \cr 
              && \cr
             &=& \sum_\X \int_{t_1}^{t_2} {\rm d} t \int_V 
                 \frac{{\partial}}{{\partial} t} \left(v^i_\X \pi^\X_i
                 \right) - \delta \int_V {\cal L}~{\rm d} V \ ,
\eea 
where the last term represents the difference in the integral before (at 
time $t_1$) and after (at time $t_2$) the displacement is put in place. \ 
If we now impose that the total energy does not change $\delta U = 0$, 
and that the velocities remain fixed, i.e.~$\partial v^i_\X / \partial t 
= 0$, then upon insertion of Eq.~(\ref{eulerx}) we find 
\bea
    \delta \int_V {\cal L}~{\rm d} V &=& \sum_\X \int_{t_1}^{t_2} {\rm d} 
    t \int_V v^i_\X \left(\overline{f}{}^\X_i - \nabla_j C^\X{}^j{}_i
    \right)~{\rm d} V \cr
    && \cr
    &\approx& \int_V \delta \xi^i \left(f_i - \nabla_j C^j{}_i
    \right)~{\rm d} V \ ,
\eea
where the second equality follows because the displacement is the same 
for each constituent, we have used Eq.~(\ref{consist}), and we have kept 
only terms linear in $\delta \xi^i$. \ Since the system is isolated, the 
net force $f_i = 0$. \ Finally, after integrating by parts, we can write
\be
    \delta \int_V {\cal L}~{\rm d} V \approx \frac{1}{2} \int_V C^{i j} 
    \left(\nabla_i \delta \xi_j + \nabla_j \delta \xi_i\right)~{\rm d} V 
    - \int_{\partial V} C^j{}_i \eta_j \delta \xi^i~{\rm d} A \ . 
    \label{work1}
\ee
The last term will vanish if we impose suitable boundary conditions on 
the control volume (or assume that the boundary is well outside the 
fluids).

We now need to establish that the deformation of the system set up by the 
displacement $\delta \xi^i$ induces a change in the metric, which in fact 
will define the $\delta$ variation in Eq.~(\ref{work1}). \ 
Consider an active coordinate transformation where the metric is pushed 
(at time $t_2$) to the new points $\overline{x}{}^i = x^i + \delta 
\xi^i$. \ Denoting the transformed metric as $\overline{g}_{i j}$, then 
ordinary tensor analysis implies
\be
    \overline{g}_{i j}(\overline{x}{}^k) = \frac{\partial 
    \overline{x}{}^k}{\partial x^i} \frac{\partial \overline{x}{}^l}
    {\partial x^j} g_{k l}(x^m + \delta \xi^m) \ .
\ee
Next we define 
\be
    \delta g_{i j} = \overline{g}_{i j}(\overline{x}^k) - g_{i j}(x^k) \ ,
\ee
and,  keeping terms up to linear order in the displacement, we find
\be
    \delta g_{i j} = \nabla_i \delta \xi_j + \nabla_j \delta \xi_i \ . 
                     \label{gchange}
\ee
Thus in Eq.~(\ref{work1}) we can rewrite the displacement on the 
right-hand side in terms of $\delta g_{i j}$. 

Next we must determine how to vary the fundamental variables $n_\X$ and 
$n^i_\X$ under the displacement. \ Recall that we are assuming that the 
number of $\X^{\rm th}$-constituent particles $N_\X$ remains fixed. \ 
This implies
\be
   \delta N_\X = \delta \int_V n_\X~{\rm d} V = 0 \implies \delta n_\X = 
   - n_\X \frac{\delta \sqrt{g}}{\sqrt{g}} \ ,  
\ee
where matrix theory (cf.~\cite{strang80:_lin_alg}) can be used to show 
\be
   \delta \sqrt{g} = \frac{1}{2} \sqrt{g} g^{i j} \delta g_{i j} \ .
\ee 
Thus, using also Eq.~(\ref{gchange}), we know how the displacement 
changes $n_\X$. \ Furthermore, we also know how the particle number 
density currents $n^i_\X$ change since $n^i_\X = n_\X v^i_\X$, but the 
velocities remain fixed, and so $\delta n^i_\X = v^i_\X \delta n_\X$. \  
Therefore, we have 
\be
    \delta n_\X = - \frac{1}{2} n_\X g^{i j} \delta g_{i j} 
                  \ , \quad
    \delta n^i_\X = - \frac{1}{2} n^i_\X g^{j k} \delta g_{j k} \ .
\ee

The small variation indicated on the left-hand side of Eq.~(\ref{work1}) 
is the difference between the integral before the displacement and the 
integral after the displacement. \ But as the integral is a function only 
of $n_\X$, $n^i_\X$, and $g_{i j}$, and the variations $\delta n_\X$ and 
$\delta n^i_\X$ induced by the displacement are equivalent to a variation 
of the metric, we see that the $\delta$ on the left-hand side of 
Eq.~(\ref{work1}) can be equally understood as a variation of the metric. 
\ Thus, Eq.~(\ref{work1}) can be reduced to
\be
   \int_V \left(C^{i j} - \frac{2}{\sqrt{g}} \frac{\partial\left(\sqrt{g} 
            {\cal L}\right)} {\partial g_{i j}}\right) 
            \delta g_{i j}~{\rm d} V = 0 \ ,
\ee
which implies
\be
   C^{i j} = \frac{2}{\sqrt{g}} \frac{\partial\left(\sqrt{g} 
            {\cal L}\right)} {\partial g_{i j}} \ .
\ee
This result should be compared to the standard formula used in general 
relativity (see, for example, \cite{mtw73}).

Finally, using the less-general internal energy given in Eq.~(\ref{eiso}) 
we find 
\be
    C^j{}_i = \Psi \delta^j{}_i + \sum_\X n^j_\X p^\X_i \ . 
\ee 
which is Eq.~(\ref{consC}) in the main body of the paper.

\section*{Appendix B}

Consider the simplest application, where each particle number is 
conserved independently and there are no forces acting on a constituent, 
i.e.~$f^\X_i = 0$. \ Then 
\be
    \partial_t{n}_\X + \nabla_i n^i_\X = 0  
\ee
and
\be
    \frac{\partial}{\partial t} \left(n_\X p^\X_i\right) + \nabla_j 
    \left(p^\X_i n^j_\X\right) = n_\X \nabla_i p^\X_0 + n^j_\X 
    \nabla_i p^\X_j \ . 
\ee 
Recall that
\be
    \nabla_i \Psi = - \sum_\X \left(n_\X \nabla_i p^\X_0 + n^j_\X 
    \nabla_i p^\X_j\right) \ .
\ee
We see that the term appearing in the summation is precisely the 
term on the right-hand-side of the above ``Euler'' equation. \ 
Hence, unless $\Psi$ takes the particular form of
\be
    \Psi = \psi^1(p^1_0,p^1_i) + \psi^2(p^2_0,p^2_i) + ... 
           + \psi^N(p^N_0,p^N_i)
         = \sum_{\sigma = 1}^N \psi^\sigma(p^\sigma_0,p^\sigma_i) \ , 
\ee 
which results in the exact differential
\be
    \nabla_i \psi^\sigma = - \left(n_\sigma \nabla_i p^\sigma_0 + 
    n^j_\sigma \nabla_i p^\sigma_j\right) 
\ee
for each constituent, such that 
\be
    n_\sigma = - \frac{\partial \psi^\sigma}{\partial p^\sigma_0} 
    \quad , \quad 
    n^i_\sigma = - \frac{\partial \psi^\sigma}{\partial p^\sigma_i} \ ,
\ee
the right-hand-side of the ``Euler'' equation will not be an exact 
differential, and so cannot be written as a total divergence.

\bibliography{biblio}

\end{document}